\documentclass[conference]{IEEEtran}
\IEEEoverridecommandlockouts
\usepackage[a4paper, total={184mm,239mm}]{geometry}
\def\BibTeX{{\rm B\kern-.05em{\sc i\kern-.025em b}\kern-.08em
    T\kern-.1667em\lower.7ex\hbox{E}\kern-.125emX}}
\usepackage{amsmath,amsfonts}
\usepackage{algorithmic}
\usepackage{algorithm}
\usepackage{array}
\usepackage[caption=false,font=normalsize,labelfont=sf,textfont=sf]{subfig}
\usepackage{textcomp}
\usepackage{stfloats}
\usepackage{url}
\usepackage{verbatim}
\usepackage{graphicx}
\usepackage{cite}

\usepackage{xcolor}
\usepackage{multirow}
\definecolor{darkgreen}{RGB}{0,120,0}

\newcommand{\squishlist}{
 \begin{list}{$\bullet$}
  { \setlength{\itemsep}{0pt}
     \setlength{\parsep}{3pt}
     \setlength{\topsep}{3pt}
     \setlength{\partopsep}{0pt}
     \setlength{\leftmargin}{1.5em}
     \setlength{\labelwidth}{1em}
     \setlength{\labelsep}{0.5em} } }

\newcommand{\squishlisttwo}{
 \begin{list}{$\bullet$}
  { \setlength{\itemsep}{0pt}
     \setlength{\parsep}{0pt}
    \setlength{\topsep}{0pt}
    \setlength{\partopsep}{0pt}
    \setlength{\leftmargin}{2em}
    \setlength{\labelwidth}{1.5em}
    \setlength{\labelsep}{0.5em} } }

\newcommand{\squishend}{
  \end{list}  }

\begin{document}


\title{Performance Implications of Multi-Chiplet Neural Processing Units on Autonomous Driving Perception}

\author{
\IEEEauthorblockN{
Mohanad Odema, Luke Chen, Hyoukjun Kwon, Mohammad Abdullah Al Faruque}

{University of California, Irvine, USA} \\

{\{modema, panwangc, hyoukjun.kwon, alfaruqu\}}@uci.edu
\vspace{-5truemm}
}



\maketitle
\pagestyle{empty}

\begin{abstract}

    We study the application of emerging chiplet-based Neural Processing Units to accelerate vehicular AI perception workloads in constrained automotive settings. The motivation stems from how chiplets technology is becoming integral to emerging vehicular architectures,
    providing a cost-effective trade-off between performance, modularity, and customization; and from perception models being the most computationally demanding workloads in a autonomous driving system.
    Using the Tesla Autopilot perception pipeline as a case study, we first breakdown its constituent models and profile their performance on different chiplet accelerators. From the insights, we propose a novel scheduling strategy to efficiently deploy perception workloads on multi-chip AI accelerators. Our experiments using a standard DNN performance simulator, MAESTRO, show our approach realizes 82\% and 2.8$\times$ increase in throughput and processing engines utilization compared to monolithic accelerator designs. 
    
\end{abstract}


\section{Introduction}

The landscape of the automotive industry is being transformed through software-defined vehicles (SDVs) that enable integrating sought-out features of Advanced Driver Assistance Systems (ADAS), autonomy and infotainment (AR/gaming) \cite{bosch2023architecture, boston2023the, fraunhofer2023chiplets, imec2023why}. 
To maintain flexibility and manage rising compute demands, emerging vehicular systems are shifting towards new cross-domain, centralized Electrical/Electronic (E/E) architectures with a few powerful processing computers and zone ECUs \cite{bosch2023architecture}, enabling easier integration of new features and updates, and allowing automakers to adopt customized chip design tailored to their needs. 
As a result, automakers like Tesla, GM Cruise, Volkswagen have entertained the adoption of customized System-on-Chips (SoCs) for their automotive compute requirements \cite{towardsAI2022tesla, fierce2022GM, Sunrise2021Volkswagen}. 
Still, as automotive AI workloads continue to evolve and rise in complexity, advancements in automotive SoC hardware also become a necessity, and that presents a considerable challenge given the long design cycle, high manufacturing costs, and Moore's law stagnation.   


Chiplets technology presents a viable solution for the automotive industry \cite{boston2023the, imec2023why, Marvell2023Marvell, understanding2023Synopsys}. Owing to chiplets' composability property, a scalable, customizable approach becomes feasible supporting the integration of individual hardware modules on the package level, potentially from different technology node generations, and enabling individual component updates to be incorporated with a faster turnaround time than monolithic SoC approaches. Even more so, recent advancements has seen Neural Processing Engines (NPUs) -- integral to autonomous driving systems -- implemented through consolidating multiple small accelerator chiplets on the package to form a scalable AI inference engine \cite{shao2019simba, cai2023gemini, tan2021nn, odema2024scar, wang2022ai, zimmer20200}, providing flexible means to control different design parameters (accelerator chips number, connection topology, and heterogeneity) \cite{odema2024scar, wang2022ai}.

\begin{figure}[!tbp]
\begin{center}
{\includegraphics[,width = 0.49\textwidth]{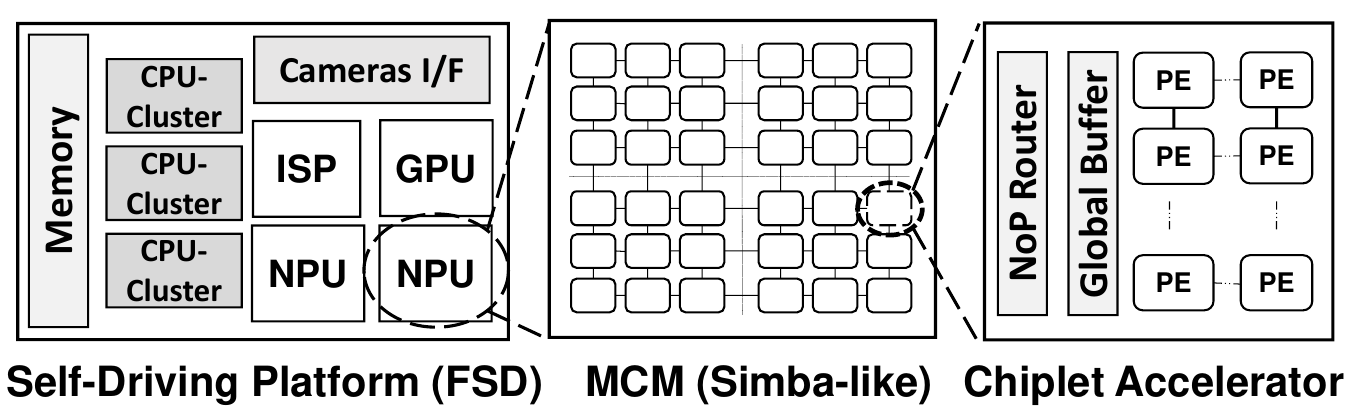}}
\end{center}
\vspace{-2ex}
\caption{A descriptive schematic showing this work's scope in adopting accelerator MCMs as NPUs in self-driving platforms (e.g., Tesla FSD).}
\label{fig:FSD_schematic}
\vspace{-4ex}
\end{figure}

Still, NPUs constructed as multi-chiplet modules (MCM) remain largely understudied in the context of automotive AI workloads despite the challenges and potential gains. On the one hand, automotive AI workloads exhibit unique execution flows characterized by intricate dependencies, concurrencies, and feature fusion nodes \cite{mullapudi2018hydranets, malawade2021sage, malawade2022hydrafusion}. On the other, AI computing kernels exhibit varying affinities towards different accelerator types \cite{kwon2021heterogeneous, odema2024scar, panopoulos2024carin, bouzidi2023map, zhang2022h2h}. Though multiple works \cite{liu2017computer, lin2018architectural, baidya2020vehicular, chen2022romanus} have investigated the architectural implications for improving automotive AI workloads' performance, the architectural implications of adopting chiplets technology remains to be studied.

This work aims to investigate such implications when an MCM AI accelerator is employed as the automotive NPU to accelerate AI perception workloads. Specifically, we follow the architectural template of Tesla's FSD chip \cite{talpes2020compute}, and simulate an industry-grade MCM inference accelerator engine -- Simba \cite{shao2019simba, zimmer20200} -- for the system's NPU (Figure \ref{fig:FSD_schematic}). For the automotive workloads, we focus on those from the compute-intensive perception pipeline \cite{lin2018architectural, teslaAIDay2021, mullapudi2018hydranets}. We implement and analyze such workloads following the Tesla Autopilot system \cite{towardsAI2022tesla} which entail HydraNets, spatio-temporal fusion, and multi-task heads (lane detection, occupancy networks). From the insights, we devise a novel scheduling methodology to enhance performance efficiency of perception workloads on the MCM-NPU. In summary, our key contributions are as follows:

\begin{itemize}

    \item We characterize the key workloads existing in a SOTA perception pipeline (Tesla Autopilot), from the early feature extraction stage till the final multi-task heads models. 

\item
We conduct a thorough performance breakdown of perception workloads to understand their execution properties and hardware acceleration affinities using the standard DNN performance simulator, MAESTRO \cite{kwon2020maestro, kwon2019understanding}.

\item From our analysis, we implement a low-cost scheduling algorithm for mapping perception workloads onto MCM-NPUs given added Network-on-Package overheads and the diversity of models \textit{within} and \textit{across} the pipeline stages.

\item We evaluate our solution against existing NPU baselines showcasing performance improvement trade-offs between utilization, energy, and pipelining latency.

\end{itemize}

\label{sec:introduction}
\section{Background and Preliminaries}

\subsection{Anatomy of a self-driving platform architecture: Tesla FSD}

We take the Tesla FSD (Full Self Driving) SoC as our reference self-driving architecture template. As illustrated in Figure \ref{fig:FSD_schematic}, the FSD integrates the following units:

\begin{itemize}
    \item \textbf{CPU clusters}. For general purpose compute. Each cluster constitutes a quad-core Cortex A72 in the Tesla FSD.
    \item \textbf{Memory}. Main memory component of the chip (e.g., LPDDR4 memory)
    \item \textbf{Cameras I/F and ISP}. High speed camera serial interface and image signal processor for image preprocessing 
    \item \textbf{NPUs}. Custom-designed hardware accelerators for efficient processing of AI workloads 
    \item \textbf{GPU}. For light-weight post-processing 
    \item \textbf{Other}. Video encoder, safety component.
\end{itemize}

For a chiplets technology variant, a system-in-package (SiP) can be constructed from this template integrating multiple smaller chiplets, each covering a subset of components and communicating over package via high-speed interconnects. Moreover, recent prototypes like Simba \cite{shao2019simba} have shown that the NPU component can also be implemented as MCMs integrating accelerator chiplets on the package level.

\begin{figure*}[!tbp]
\begin{center}
{\includegraphics[,width = 0.97\textwidth, height=0.38\textwidth]{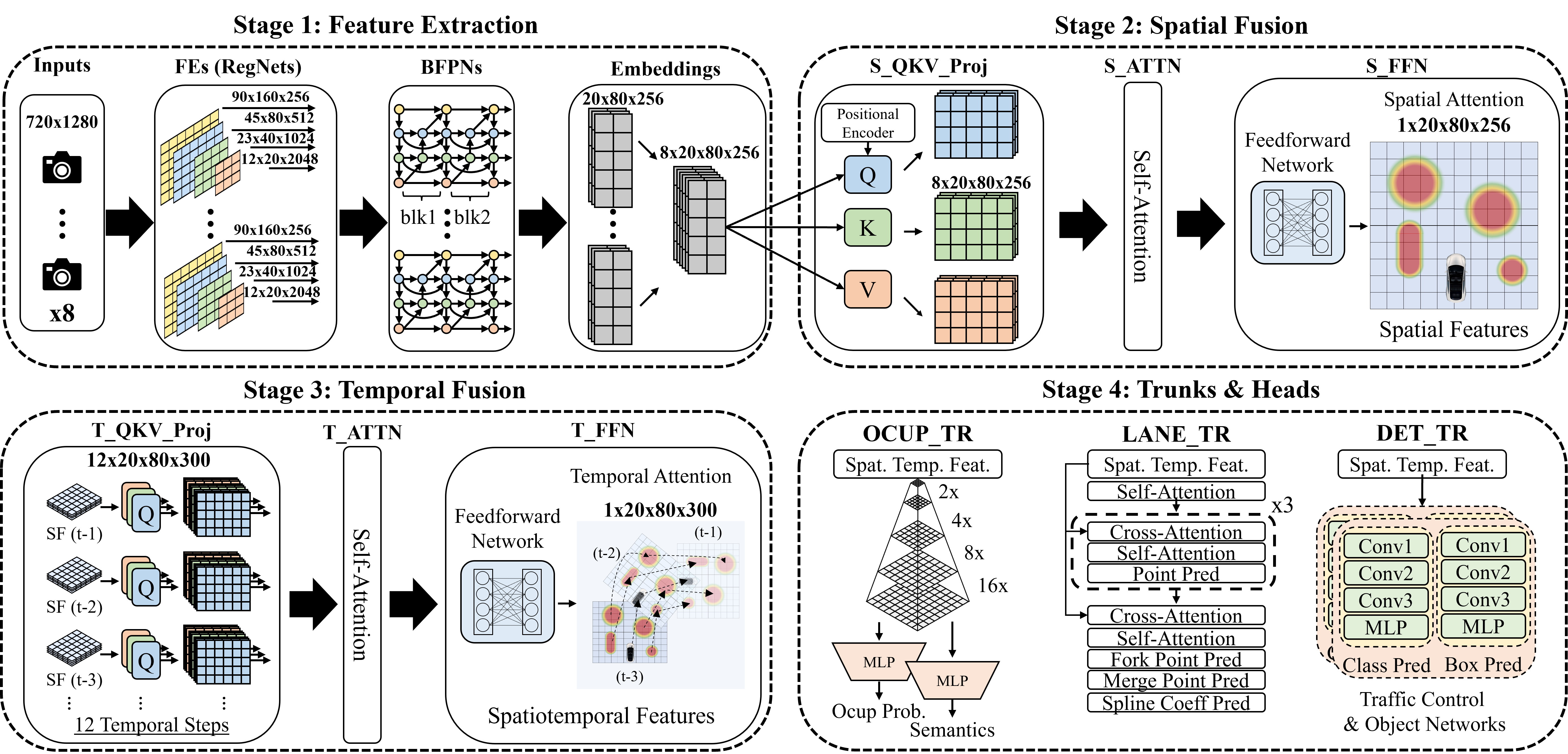}}
\end{center}
\vspace{-2ex}
\caption{The four-stage perception pipeline based on the HydraNet architecture \cite{mullapudi2018hydranets} by Tesla Autopilot system \cite{autonomous2021how} whose feature dims and models are displayed.}
\label{fig:av_percep_pipe}
\vspace{-2ex}
\end{figure*}

\subsection{Autonomous Driving System Pipeline} \label{subsec:autopilot}

Autonomous driving systems (ADS) map sensory input data to vehicle control outputs through 3 stages: \textit{perception} for contextual understanding and objects tracking in the driving scene; \textit{planner} for trajectory paths generation; \textit{control} for providing the necessary control outputs. Perception represents the most compute intensive stage \cite{lin2018architectural, malawade2021sage}, and supports multiple driving tasks (detection, lane prediction). Recent perception modules have seen the adoption of the HydraNets architecture \cite{mullapudi2018hydranets}, with a \textit{shared backbone} for extracting low-level, common features, and \textit{specialized heads} for driving tasks specialization. In Figure \ref{fig:av_percep_pipe}, we illustrate the HydraNet architecture from the Tesla Autopilot system\cite{autonomous2021how} discussed in the following:

\textbf{Sensor inputs.} Tesla Autopilot perception relies on collecting images from 8 installed cameras. Typically images can be 720p resolution at around 30 FPS \cite{lin2018architectural}.

\textbf{Stage 1: Feature Extraction (FE+BFPN).} Each raw input image is pre-processed and passed through a feature extractor (FE) followed by a bidirectional feature pyramid network (BFPN) \cite{tan2020efficientdet}, to generate multi-scale feature representations.
Following \cite{teslaAIDay2021}, the FE can be a ResNet18 architecture with 4 multiscale features (90x160x256, 45x80x512, 23x40x1024, and 12x20x2048) generated throughout its ResNet blocks, which are then passed through 2 Bi-directional FPN blocks (BFPN). At the final output, multiscale features are concatenated from each FE+BFPN pipeline forming the 8x20x80x256 outputs.

\textbf{Stage 2: Multi-cam Spatial Fusion (S\_FUSE).} Generated embeddings from each camera are fused onto a combined spatial representation in vector space (2D/3D grid map). A transformer with multi-attention head is employed, computing attention scores between every grid cell and detected features from each camera.
The attention module \cite{vaswani2017attention} comprises: \textit{QKV projection} to generate Query (Q), Key (K), and Value (V) vectors; \textit{Attention} with two matrix multiplications for (QK$^T$)$\cdot$V; \textit{FFN} (Feed-forward Network) comprising two linear layers. For a 20$\times$80 2D grid \cite{autonomous2021how,teslaAIDay2021}, the output becomes a fused projection of the 8 camera features onto a 1x20x80x256 grid.

\textbf{Stage 3: Temporal Fusion (T\_FUSE).} The 1x20x80x256 spatial representation is fed into a video queue to be fused with a feature queue of N previous representations to accommodate necessary temporal features (e.g., seen signs or lane markings), and telemetry information. Another attention module can be used with N=12 for temporal fusion \cite{autonomous2021how, teslaAIDay2021}, each undergoing QKV projection, attention, and FFN transformation till the final fused spatio-temporal 1x20x80x300 representation. 

\textbf{Stage 4: Trunks and Heads (TR).} The 1x20x80x300 representation is fed to branched trunks and heads including: 
\squishlist
    {\item \textbf{Occupancy network} predicting the grid's continuous occupancy probability and continuous semantics, involving 4 spatial deconvolution layers with 16x upscaling}
    {\item \textbf{Lane prediction} involving a combination of self-attention and cross-attention layers repeated 3 times for 3 levels of point predictions, and having 3 classifier predictors }
    {\item \textbf{Object detection} Multiple detection heads (traffic, vehicle, pedestrians) are supported. Each detector head entails separate class and box prediction networks using a sequence of 3 convolution layers and fully connected layer. }

\squishend

\label{sec:background}
\section{Analysis of Perception Module Workloads}
\label{sec:analysis}

To derive design insights for MCM-based NPUs, We first analyze the perception workloads' performance on various accelerator architectures from Section \ref{subsec:autopilot} using the analytical DNN cost model simulator, MAESTRO \cite{kwon2020maestro, kwon2019understanding} (known to have achieved 96\% accuracy compared to RTL simulations). We consider the following:
\squishlist
{\item We set the number of PEs in each accelerator to 256 PEs. That way, when accelerators are assimilated as chiplets into a 6$\times$6 MCM like Simba \cite{shao2019simba}, a total of 9,216 PEs will be available equivalent to that of the original Tesla NPU \cite{talpes2020compute}. We use the same operating frequency at 2 GHz \cite{talpes2020compute}.} 
{\item We analyze the workloads on weight stationary (WS) and output stationary (OS) accelerator types -- like NVDLA \cite{nvdla} and Shidiannao \cite{du2015shidiannao}) -- given their proven superiority over other accelerator types \cite{kwon2021heterogeneous, kim2023dream, odema2024scar}.}
\squishend

\begin{figure}[!tbp]
\begin{center}
{\includegraphics[,width = 0.49\textwidth]{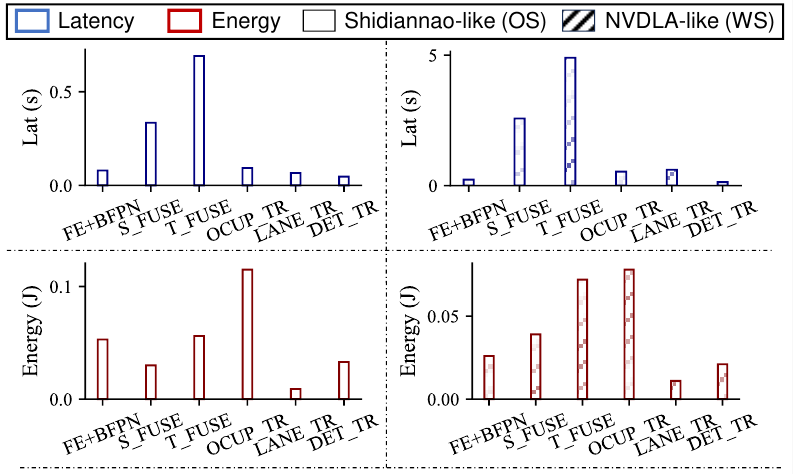}}
\end{center}
\vspace{-2.6ex}
    \caption{Breakdown of latency (\textit{top}) and energy consumption (\textit{bottom}) per perception component across Shidiannao- (\textit{left}) and NVDLA-like (\textit{right}) accelerators  using MAESTRO for a single 256 accelerator chiplet.}
\label{fig:breakdown}
\vspace{-2ex}
\end{figure}

\subsection{Coarse-grained Performance Analysis}

In Figure \ref{fig:breakdown}, we breakdown the contributions of the perception stage models in terms of latency and energy on NVDLA- and Shidiannao-like accelerators and deduce the following:

\textbf{OS dataflow suits latency-critical workloads.} Across all workloads, the Shidiannao OS dataflow offers \textbf{6.85$\times$} speedups over its WS counterparts, making it the prime candidate for latency critical automotive workloads \cite{lin2018architectural}.

\textbf{WS offers energy efficiency opportunities.} On average, the WS dataflows can lead to \textbf{1.2$\times$ }energy efficiency gains compared to OS. The gains are even more significant (\textbf{1.55$\times$}) if we omit S\_FUSE and T\_FUSE from our analysis, which are more affine towards OS dataflow. This opens the door for potential heterogeneous integration (mix of OS and WS chiplets) to realize interesting performance trade-offs.

\textbf{Fusion Modules are computational bottlenecks.} The S\_FUSE and T\_FUSE modules constitute respective \textbf{25\%-28\%} and \textbf{52\%-54\%} of the overall perception module latency under both dataflow scenarios. This is attributed to the feature aggregation from multiple sources (8 cameras and 12 temporal frames) onto a shared projection space (200$\times$80$\times$256 grid). 

\textbf{FE+BFPN Scaling.} Evaluations for the FE+BFPN in Figure \ref{fig:breakdown} are for a single camera and to be multiplied by the 8 cameras.

\begin{figure}[!tbp]
\begin{center}
{\includegraphics[,width = 0.49\textwidth]{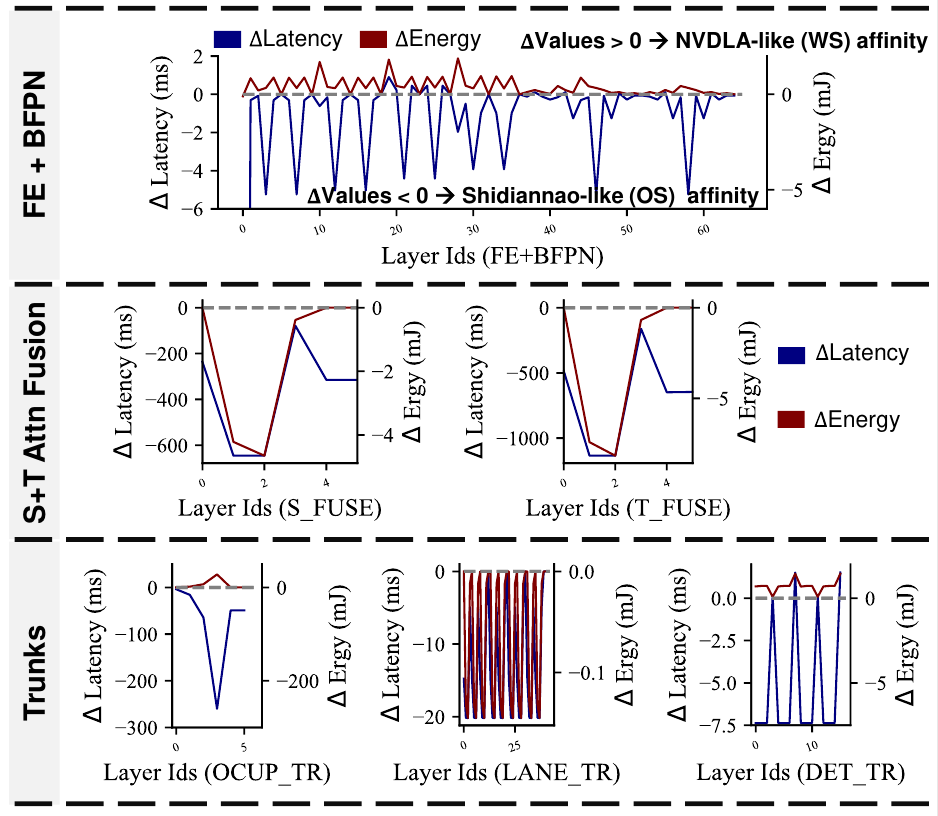}}
\end{center}
\vspace{-2.6ex}
    \caption{Affinities of the feature extractors (\textit{top}), spatio-temporal attention fusion \textit{(mid)}, and trunks (bot) towards Shidiannao- and NVDLA-like accelerators. 
    $\Delta$Value$<$0 implies Shidiannao-like affinity and the opposite for NVDLA-like. 
    }
\label{fig:affinities}
\vspace{-3ex}
\end{figure}

\subsection{Fine-grained Performance Analysis}

We analyze the individual layers affinities towards OS and WS from the various models in Figure \ref{fig:affinities}, where we compare the difference in values between the OS and WS dataflows, $\Delta Value$ = $Value_{OS}$ - $Value_{WS}$, for latency and energy (\textit{-ve values imply OS affinity and +ve values imply WS affinity}).

\textbf{FE+BFPN tradeoffs.} The FE+BFPN stage exhibits a direct trade-off between latency and energy across all layers. Yet, the non-uniform distribution of performance gains, the complex dependencies of the FE and BFPN networks, and the concurrent execution requirement for 8 FE+BFPN models impose strict requirements on accelerator chiplets assignment process.

\textbf{S\_FUSE and T\_FUSE tradeoffs.} The negative evaluations of $\Delta$Latency and $\Delta$Energy across all layers indicate a strong affinity towards the OS dataflow for the fusion modules. We also observe significant performance bottlenecks exist at the self-attention layers (QKV calculations at layer ids 1 and 2).

\textbf{Trunks tradeoffs.} The diversity of trunk models lead to varying affinities: the lane prediction trunk is completely skewed towards the OS dataflow (due to attention modules), unlike other trunks which can provide exploitable trade-offs.

\subsection{Design Insights.}

Based on our analysis, we derive the following insights onto scheduling perception workloads onto MCM-based NPUs:

\begin{itemize}

    \item Any scheduling configuration is to target maintaining a consistent throughput (pipelining latency) across the various perception stages for streamlined input processing.

    \item Heterogeneous chiplets integration can be supported as long as the latency constraint is not violated. Given the dominance of the OS dataflow with regards to execution latency, we specify the pipelining latency of the OS dataflow as the reference latency constraint.

    \item Any optimizations for the FE+BFPN stage must be applied for the 8 concurrent models for simultaneous execution.
    
    \item The computational bottleneck of the fusion layers are confined within a small number of layers (see Figure \ref{fig:affinities}), making them targets for parallelization optimizations.

    \item Heterogeneous chiplets integration can be particularly beneficial for the diverse trunk workloads to elevate efficiency.

\end{itemize}
\section{Proposed Scheduling Methodology}
\label{sec:scheduling_framework}

We propose to schedule the perception workloads on MCM-NPU through a throughput matching mechanism as follows:

\begin{enumerate}
    \item Specify initial execution latency estimates and chiplet allocation per layer in each stage
    \item Allocate chiplet resources across the various workload stages based on the latency estimates
    \item Alleviate bottleneck stages impact via data sharding 
    \item Update latency estimates and chiplet assignments 
    \item Repeat steps 2-4 until pipelining latencies are matched or no further sharding is possible
\end{enumerate}

In Algorithm 1, we achieve this sequence through a nested greedy algorithm whose outer loop alleviates bottlenecks across stages, and inner loop alleviates bottlenecks across layers in the bottleneck stages. \textit{\textbf{Line 2}} demonstrates how chiplets are initially allocated to each stage based on a function of the model execution latencies, the number of workloads, and the number of models per each stage workload. We initially adopt uniform partitioning of chiplet resources across the 4 stages, where each is assigned a separate quadrant of chiplets. The reasons being: (i) having 4 distinctive perception stages; (ii) having 8 processing parallel model instances of the FE+BFPN; (iii) the latency bottlenecks being in the fusion modules with small number of layers; (iv) the diversity of the trunk models.

\begin{figure}[!tbp]
\begin{center}
{\includegraphics[,width = 0.48\textwidth]{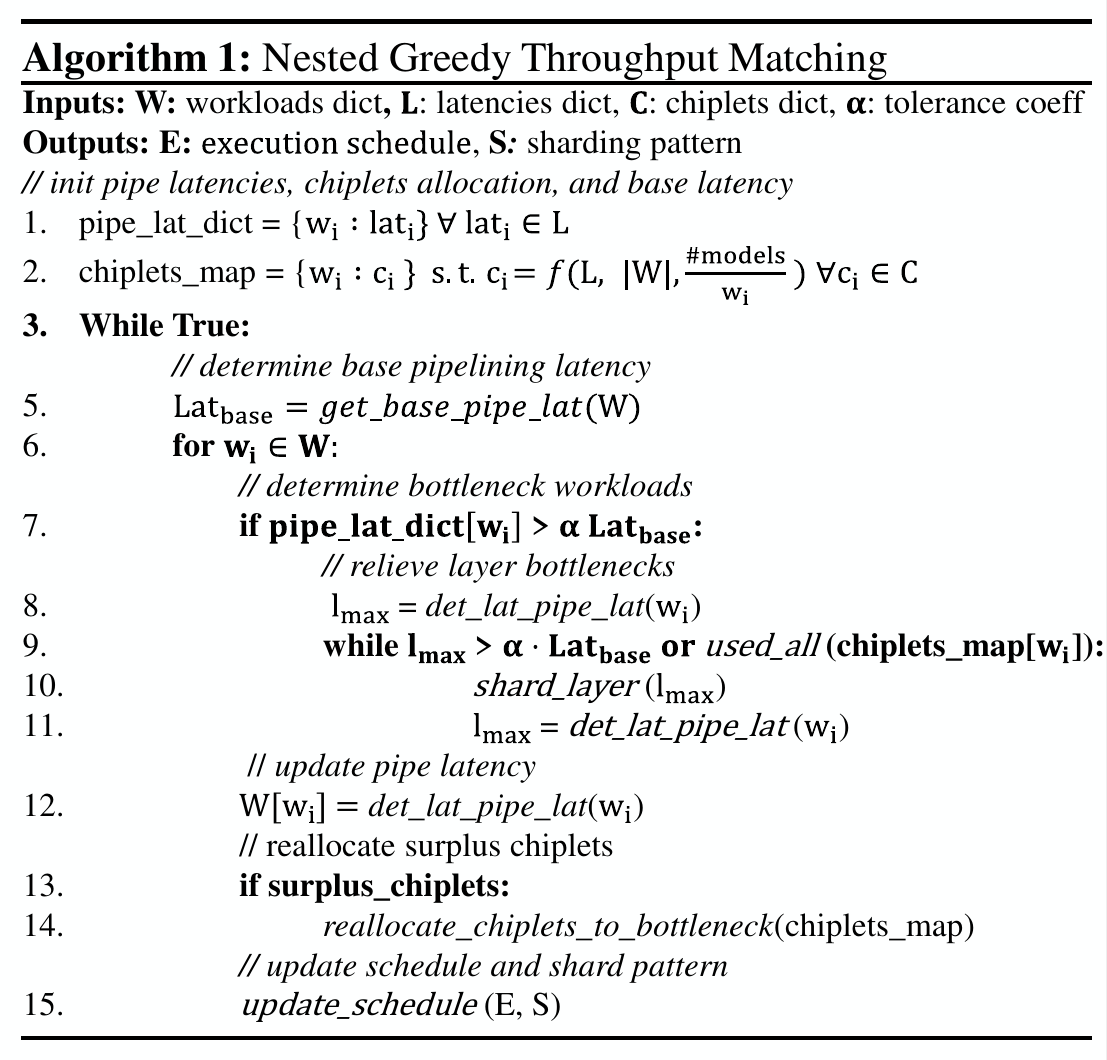}}
\end{center}
\vspace{-2.6ex}
\label{fig:alg}
\end{figure}

\begin{figure}[!tbp]
\begin{center}
{\includegraphics[,width = 0.48\textwidth]{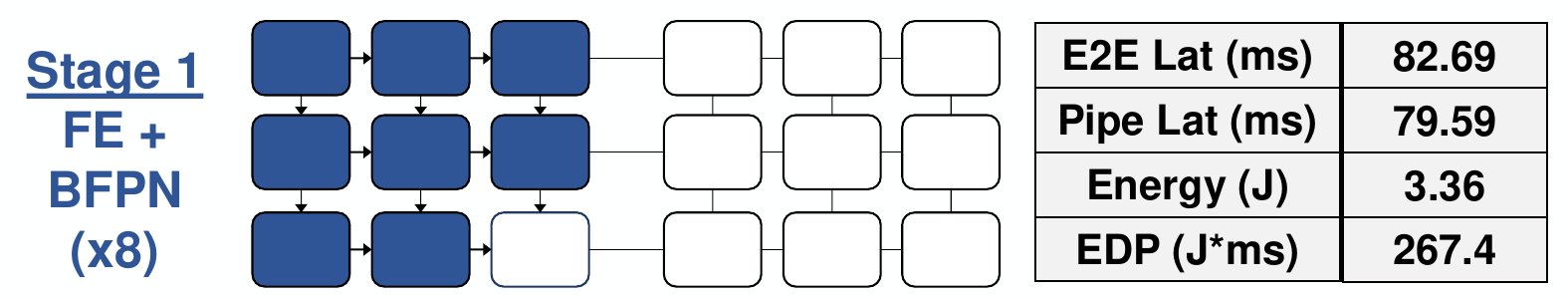}}
\end{center}
\vspace{-2.6ex}
    \caption{The 8 FE+BFPN models mapping onto the MCM's first quadrant.}
\label{fig:FE_BFPN}
\vspace{-2ex}
\end{figure}

\begin{figure}[!tbp]
\begin{center}
{\includegraphics[,width = 0.48\textwidth]{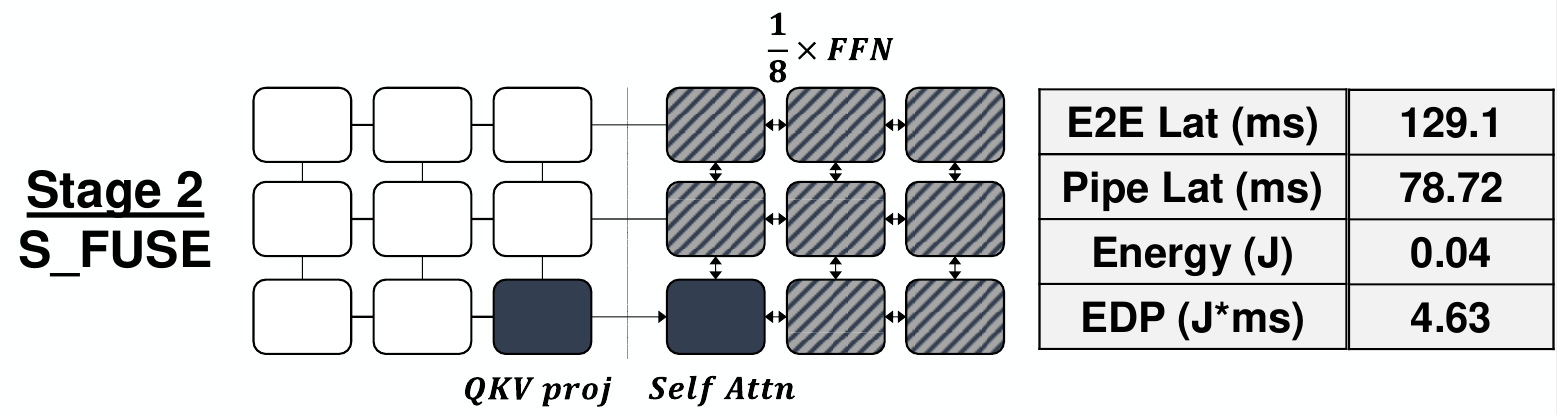}}
\end{center}
\vspace{-2.6ex}
    \caption{Spatial Fusion (S\_FUSE) mapping onto the MCM. The FFN is sharded accross 8 chiplets to maintain the pipelining latency as FE+BFPN. }
\label{fig:S_FUSE}
\vspace{-2ex}
\end{figure}

\subsection{Choosing the FE+BFPN for the base pipelining latency }

Given 8 model instances of the FE+BFPN stage where each possesing a compute latency around $\frac{1}{8}$ of the T\_FUSE bottleneck (Figure \ref{fig:breakdown}), at least 8 chiplets need to be initially allocated to each model to maintain concurrent execution. 
From the latency contribution breakdown in Figure \ref{fig:breakdown} across the 4 stages, assigning additional resources to the FE+BFPN would lead to suboptimal solution since it leaves just above half of the total number of chiplets for processing $>$90\% of the overall workloads. 
Therefore, it becomes reasonable to specify the initial base pipelining latency to that of the FE+BFPNs models ($Lat_{base}$=82.7 ms) as shown in Figure \ref{fig:FE_BFPN}.

\subsection{Alleviating S\_FUSE and T\_FUSE bottlenecks} \label{subsec:mappings}

We alleviate the attention fusion bottlenecks through recursive sharding until the stages' latency equates that of $Lat_{base}$.

\textbf{S\_FUSE Bottleneck.} This attention module operates on 8 feature input sets from the FE+BFPN stage outputs. Given 200$\times$80$\times$256 attention grid dimensions, S\_FUSE stage incurs 78.7 ms, 20.5 ms, and 236 ms latency overheads for the QKV projection, self-attention, and Feed-Forward (FFN) layers when each is processed on an individual chiplet. Accordingly, the spatial FFN layer is sharded in a four-folded manner, replicating the FFN weights on 4 chiplets, each processing features from two FE+BFPNs. This brings FFN latency close to the QKV projection (78.7 ms) and $Lat_{base}$ (82.7 ms). As S\_FUSE is still left with 4 chiplets (one is surplus from the FE+BFPN stage), an additional sharding step can be performed, leading to the final configuration shown in Figure \ref{fig:S_FUSE}.

\begin{figure}[!tbp]
\begin{center}
{\includegraphics[,width = 0.48\textwidth]{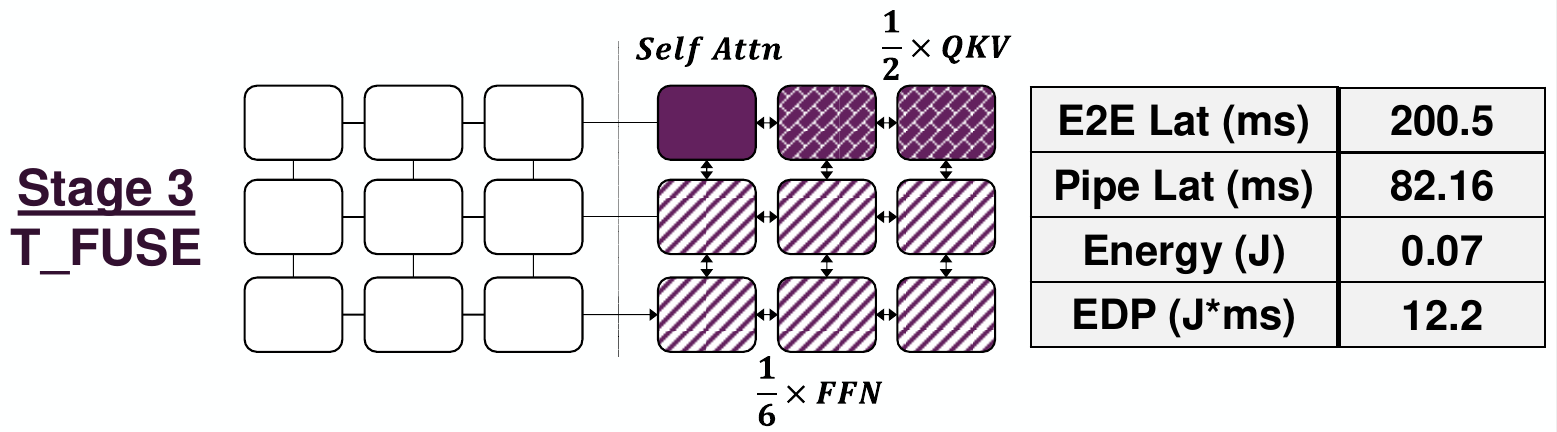}}
\end{center}
\vspace{-2.6ex}
    \caption{Temporal Fusion (T\_FUSE) mapping onto the MCM's third quadrant. QKV projection and FFN are divided across 2 and 6 chiplets, respectively. }
\label{fig:T_FUSE}
\vspace{-2ex}
\end{figure}

\textbf{T\_FUSE Bottleneck.} 
Prior to sharding the 12 frames, the T\_FUSE took 165.6 ms, 36.4 ms, and 490.2 ms for the QKV projection, self-attention, and FFN blocks, respectively. Our algorithm in this case involves two inner loop iterations, first distributing the FFN block workloads across 6 chiplets, leading FFN layer pipeline latency to drop down to 81.7 ms close to $Lat_{base}$. Then, the QKV projection is partitioned across two chiplets dropping the layer pipelining latency to 78.7 ms. All 9 chiplets in the quadrant are utilized as shown in Figure \ref{fig:T_FUSE}.

\begin{figure}[!tbp]
\begin{center}
{\includegraphics[,width = 0.48\textwidth]{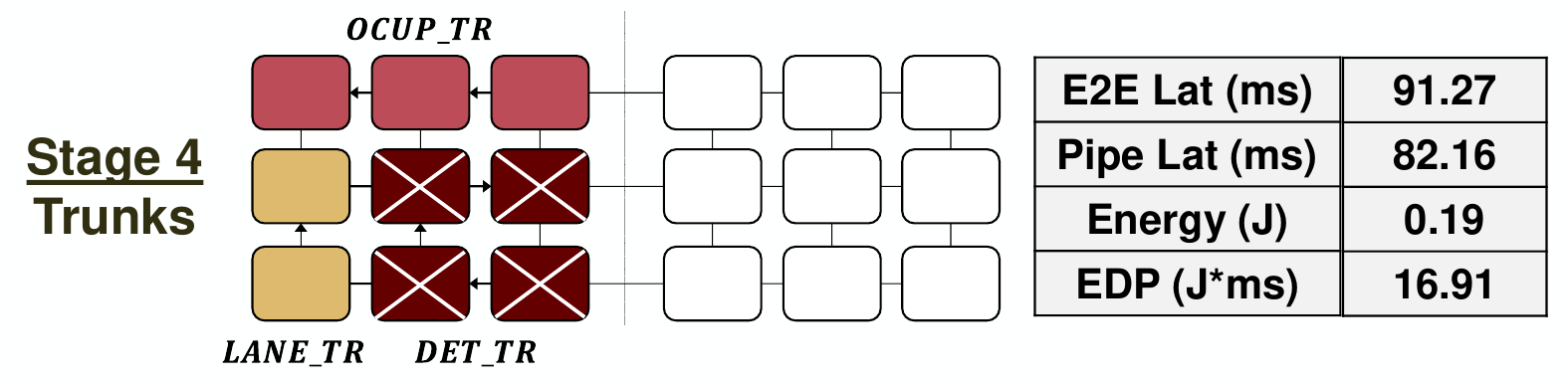}}
\end{center}
\vspace{-2.6ex}
    \caption{Mapping of Trunks onto the fourth quadrant. Chiplets marked with the cross indicate potential placement as WS style accelerators. }
\label{fig:TRUNKS}
\vspace{-2ex}
\end{figure}

\subsection{Design Space Exploration for the Trunks stage}

Given the diversity of the trunk models and their customization, we employ a design space exploration for this stage to manage the mapping of workloads onto chiplets. Additionally, we consider heterogeneous integration options based on the analysis in Figure \ref{fig:affinities} to leverage potential energy efficiency gains. In particular, we consider two configurations, \textbf{Het(2)} and \textbf{Het(4)}, which integrate 2 and 4 WS chiplets within the OS-dominated 3$\times$3 chiplets quadrant, respectively. We specify the following scoring function for our search:
\begin{equation*}
    Score(config) = 
    \begin{cases}
        - \infty, \;\;\;\;\;\;\;\;\;\;\;\text{if $\exists L_{max}$(chiplet) $> L_{cstr}$ } \\
        -1 \times EDP, \;\; otherwise
    \end{cases}
\end{equation*}

where a scheduling configuration ($config$) is only considered on the basis of its Energy-Delay Product evaluation as long as the pipeline latency ($L_{cstr})$ is not violated by any chiplet. 
Given the relatively small size of the search space (9 chiplets, 3 models, and $<$ 100 layers), we perform a brute force search with the results provided in Table \ref{tab:DSE} at \textbf{$L_{cstr}$ = 85 ms}. 

\begin{table}[t]
    \centering
    \renewcommand{\arraystretch}{1.2}
    \caption{Results of heterogeneous integration for the MCM trunks relative to the OS only configuration. $L_{cstr}$ is set to 85 ms. $\Delta$ means the \% change between hetero. and OS configurations.}
    \vspace{-2ex}
    \begin{tabular}{ l | c c | c  c | c c }
    \hline
    Metric & OS & WS & Het(2) & Het(4) & $\Delta$(2) & $\Delta$(4) \\
    \hline
    \textbf{E2E Lat(ms)} & 91.2 & 605.7 & 91.3 & 91.3 & +0.1\% & +0.1\% \\
    \textbf{Pipe Lat(ms)} & 87.9 & 605.7 & 71.7 & 71.7 & -18.4\% & -18.4\% \\
    \textbf{Energy(J)} & 0.185 & 0.139 & 0.183 & 0.174 & \textbf{-1.1\%} & \textbf{-6.2\%} \\
    \textbf{EDP(ms*J)} & 16.89 & 59.35 & 14.38 & 15.1 & \textbf{-17.4 \%} & \textbf{-12.0\%}\\
    \hline
    \end{tabular}
    \label{tab:DSE}
\end{table}

From the table, we can see that the heterogeneous configurations, \textbf{Het(2)} and \textbf{Het(4)} realize performance efficiency improvements in energy and EDP compared to the \textbf{OS} only configuration, \textbf{achieving average \textbf{1.1\%} and \textbf{6.2\%} reductions in raw energy; \textbf{17.4\%} and \textbf{12.0\%} reductions in EDP, respectively.} 
Upon careful inspection, we found the WS chiplets are predominantly assigned to the DET\_TR layers, leading DET\_TR independently to achieve a 35\% reduction in energy.

\subsection{NoP Cost Modeling}

We model the NoP data movement overheads using a cost model built around MAESTRO \cite{kwon2020maestro, kwon2019understanding} and the microarchitecture parameters from Simba \cite{shao2019simba} scaled to 28 nm as follows:
\begin{itemize}
    \item NoP Interconnect BW: 100 GB/s/chiplet
    \item NoP Interconnect Lat: 35 ns/hop
    \item NoP Interconnect Ergy: 2.04 pJ/bit
\end{itemize}
Transmission latency is computed as a function of feature map size over the NoP bandwidth, multiplied by the number of hops from source to destination. Whereas transmission energy is the multiplication of feature map size, interconnect bit transmission energy, and the number of hops.  

\begin{figure}[!tbp]
\begin{center}
{\includegraphics[,width = 0.48\textwidth]{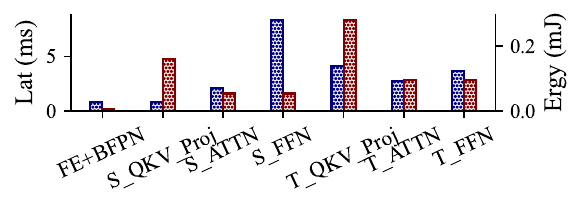}}
\end{center}
\vspace{-4ex}
    \caption{The NoP data movement costs in terms of latency (\textit{blue}) and energy (\textit{maroon}) throughout the first 3 stages of the perception pipeline.}
\label{fig:NoP}
\vspace{-2ex}
\end{figure}

We analyze the NoP data movement overheads in Figure \ref{fig:NoP} across the different layer workloads following the scheduling configuration of the previous subsections and make the following observations: (i) Large feature map outputs (as from QKV\_Proj layers) have high transmission costs (latency and energy) and are placed in close proximity to their destinations to limit overheads (Figures \ref{fig:S_FUSE} and \ref{fig:T_FUSE}); (ii) The gathering of sharded outputs from multiple far nodes (S\_FFN) can cause rise in NoP traffic and the corresponding latency. (iii) Most importantly, the overall perception latency bottleneck does not lie in NoP transmission overheads as they are at least two orders of magnitude less than the computational costs in Figure \ref{fig:breakdown}. 

\section{Evaluation}

We evaluate our MCM scheduling solution against baselines with the same number of PEs:
\begin{itemize}
    \item A single accelerator chiplet with 9,216 PEs (Regular)
    \item Two accelerator chiplets with 4,608 PEs each 
    \item Four accelerator chiplets with 2,304 PEs each
\end{itemize}

\begin{table}[t]
    \footnotesize
    \centering
    \renewcommand{\arraystretch}{1.2}
    \caption{Performance evaluation for various chiplet arrangements for the same total number of PEs (9,216). 36$\times$256 is Simba-like \cite{shao2019simba}}
    \vspace{-2ex}
    \begin{tabular}{l | l | c | c | c | c }
    \hline
    \hline
    Pipeline & Metric & 1x9216 & 2x4608 & 4x2304 & 36x256 \\
    \hline
    \multirow{4}{*}{Stagewise} &
    \textbf{E2E Lat(s)} & 1.8 & 1.8 & 1.8 & 0.5 \\
    & \textbf{Pipe Lat(s)} & 1.8 & 0.7 & 0.67 & 0.09 \\
    & \textbf{Energy(J)}  & 0.64 & 0.69 & 0.65 & 0.71 \\
    & \textbf{EDP(ms*J)} & 274 & 283 & 273 & 69 \\
    & \textbf{Utilization(\%)}  & 19.11 & 25.39 & 31.13 & 54.19 \\
    \hline
    \hline
    \multirow{4}{*}{Layerwise} &
    \textbf{E2E Lat(ms)} & 1.8 & 1.5 & 1.3 & 0.5 \\
    & \textbf{Pipe Lat(ms)} & 1.8 & 0.5 & 0.5 & 0.09 \\
    & \textbf{Energy(J)}  & 0.64 & 0.64 & 0.65 & 0.71 \\
    & \textbf{EDP(ms*J)}  & 274 & 141 & 86 & 69 \\
    & \textbf{Utilization(\%)}  & 19.11 & 26.20 & 31.86 & 54.19 \\
    \hline
    \hline
    \end{tabular}
    \label{tab:comparison}
\end{table}

We consider two pipelining schemes for the evaluation: (1) \textbf{stagewise}, and (2) \textbf{layerwise}.
We also analyze  how our solution can scale to two NPUs (2$\times$6$\times$6 Simba MCMs for a total of 72 chiplets).
Unless otherwise stated, we focus our analysis on the multi-chiplet NPU with OS only dataflow. We perform comparisons on the first 3 bottleneck stages of the perception pipeline. A separate ablation study is performed for the trunks.

\subsection{Comparison against Baselines}

We show the comparison against the baselines in Table \ref{tab:comparison}. Though the 6$\times$6 solution achieves the best pipelining latency scores with 87 ms, it incurs additional energy consumption from the overhead associated with the NoP data transmission, leading it to incur a 10.9\% increase in energy consumption compared to the single chiplet solution. Still, the 6$\times$6 solution achieves the lowest EDP of 69 J*ms as a result of low end-to-end latency from the cross-chiplet sharding. 
In terms of PEs utilization, the 6$\times$6 solution achieves an average of 54.19\% utilization across all chiplets' PEs, constituting a 2.8$\times$, 2.1$\times$, and 1.7$\times$ increase over the three respective baselines.

\subsection{Scaling to 2 Multi-chiplet NPUs}

\begin{figure}[!tbp]
\begin{center}
{\includegraphics[,width = 0.48\textwidth]{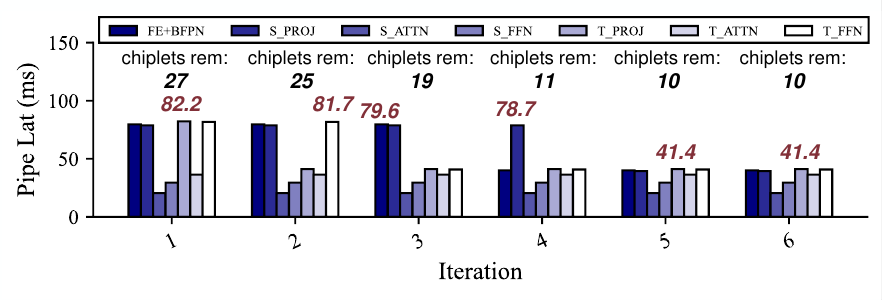}}
\end{center}
\vspace{-2.6ex}
    \caption{Algorithm 1 steps when alleviating pipelining latency when two Simba NPUs are active. Numbers in (\textit{red}) indicate corresponding pipelining latency.
    }
\vspace{-2ex}
\label{fig:scale}
\end{figure}

We explore how our solution scales to 72 chiplets if the two NPUs on the FSD are active and processing the same workloads (recall Fig \ref{fig:FSD_schematic}). We assume in this analysis the number of trunks is doubled (assigned 2$\times$9 chiplets), and that they incur a fixed performance overhead not becoming the the latency bottleneck.

In Figure \ref{fig:scale}, we show how our algorithm progresses to reduce the pipelining latency and assigns chiplet resources. First, our algorithm extends the sharding of the temporal projection layer T\_QKV from 2 to 4, reducing its pipelining latency to 41.1 ms. Then, T\_FFN layer is assigned an additional 6 chiplets to bring its latency down to 40.8 ms. At this point, the sharding is exhausted for the T\_FFN as each temporal frame is processed independently on a separate chiplet. The FE+BFPN layer is then partitioned into two pipelining stages at the fourth convolutional ResNet-18 block, yielding two equivalent partitions with pipelining latency of 40.04 ms. Lastly, S\_PROJ is divided across two chiplets to yield a latency at 39.4 ms. The final pipelining latency is 41.1 ms, almost 2$\times$ that of the 36 chiplet case.

\subsection{Ablation on Optimizing Trunk models Design}

\begin{table}[t]
    \centering
    \setlength{\tabcolsep}{1.2pt}
    \renewcommand{\arraystretch}{1.2}
    \caption{Analysis on the input scaling effects on latency for OCUP\_TR}
    \vspace{-2ex}
    \begin{tabular}{ l | c | c | c | c }
    \hline
    Upsampling Factor & [2X,2Y] & [4X,4Y] & [8X,8Y] & [16X,16Y] \\
    \hline
    \textbf{E2E Lat(ms)} & 0.97 & 4.97 (4.10x$\uparrow$) & 21.16 (20.72x$\uparrow$) & 86.29 (87.59x$\uparrow$) \\
    \textbf{Pipe Lat(ms)} & 0.97 & 3.99 (3.11x$\uparrow$) & 16.18 (15.64x$\uparrow$) & 65.13 (66.00x$\uparrow$) \\
    \hline
    \end{tabular}
    \label{tab:occup}
\end{table}

We performed additional ablation studies on the trunk models to pinpoint their performance bottlenecks.

\textbf{Occupancy Trunk.} Deconvolution operations are the bottleneck of the occupancy trunk. In Table \ref{tab:occup}, we showcase the non-linear increase in latency incurred in the occupancy network with each added upsampling layer, with the final upsampling layer contributing the most to the overall latency ($\sim$75\%). This underpins an exploitable trade-off between resolution granularity and performance efficiency.

\begin{figure}[!tbp]
\begin{center}
{\includegraphics[,width = 0.45\textwidth]{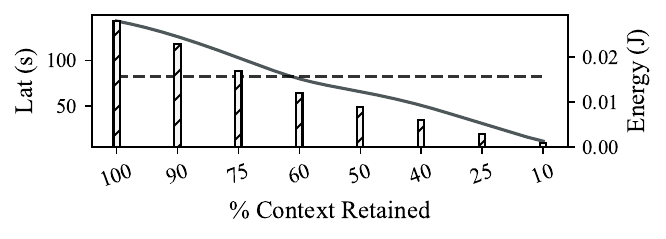}}
\end{center}
\vspace{-2.6ex}
    \caption{Lane trunk latency (\textit{lineplot}) and energy (\textit{barplot}) under context-aware computing. Dashed line indicates pipelining latency threshold at 82 ms. }
\label{fig:lane}
\vspace{-2ex}
\end{figure}

\textbf{Lane Prediction Trunk. } 
In the Tesla Autopilot System, the lane prediction network adopts a form of context-aware computing for efficiency, where rather than running the lane prediction algorithm for every region, processing is only performed for relevant regions in the grid \cite{teslaAIDay2022}. In Figure \ref{fig:lane}, we show a similar pattern in our implementation where processing across all regions leads latency to exceed the 82 ms pipelining latency. Around 60\% computing satisfies the latency constraint. 

\label{sec:eval}

\section{Related Works}

\subsection{Multi-chiplet Modules}

Accelerating AI workloads through a chiplets system involves combining on package a host system with an acceleration platform which can be based on GPU \cite{arunkumar2017mcm,  grace_hopper}, FPGA \cite{hwang2020centaur}, or NPUs \cite{shao2019simba, zimmer20200, cai2023gemini}. 
More sophisticated architectures can entail the acceleration platform comprising multiple smaller modules consolidated on package to form a multi-chiplet module (MCM) \cite{arunkumar2017mcm, shao2019simba, odema2024scar, zimmer20200, cai2023gemini, tan2021nn}.

\subsection{Autonomous Driving Systems Efficiency}

The works in
\cite{malawade2021sage, chen2022romanus, odema2022testudo, liu2019edge} study how distributed scheduling of ADS perception tasks across multiple platforms elevates efficiency.
Other works in \cite{lin2018architectural, liu2017computer} explore how different hardware architectures and platforms (GPUs, FPGAs, ASICs) influence efficiency in the autonomous driving SoC.

\label{sec:related_works}
\section{Conclusion}

We have studied the performance implications from adopting chiplet-based NPUs to accelerate perception AI workloads through the Tesla Autopilot use case. Our findings demonstrate that the combination of throughput matching algorithm and heterogeneous integration offer desirable performance trade-offs from the multi-chiplet module despite the added NoP costs, motivating further studies on adopting forthcoming chiplet-based NPU architectures in the automotive setting.

\label{sec:conclusion}

\bibliographystyle{IEEEtran}
\bibliography{ref}


\end{document}